\documentclass[manuscript,screen,nonacm]{acmart}

\AtBeginDocument{%
  }

\setcopyright{none}
\renewcommand\footnotetextcopyrightpermission[1]{}
\title{Reading the Same Data Differently: Interpretive Labor Across System Boundaries in Electronic Monitoring}

\author{Yibo Meng*}
\affiliation{
  \institution{Tsinghua University}
  \country{China}
}
\email{mengyb22@tsinghua.org.cn}

\author{Bingyi Liu}
\affiliation{
  \institution{University of Michigan, Ann Arbor}
  \country{USA}
}
\email{bingyi@umich.edu}

\author{Zhiqi Gao}
\affiliation{%
  \institution{Nankai University}
  \country{China}
}
\email{zhiqigao@link.cuhk.edu.cn}

\author{Shuai Ma}
\affiliation{%
  \institution{Aalto University}
  \country{Finland}
}
\email{shuai.ma@aalto.fi}

\author{Hongyu Zhou}
\affiliation{
  \institution{University of Cambridge}
  \country{UK}
}
\email{hz548@cam.ac.uk}


\begin{document}

\begin{abstract}
Electronic monitoring (EM) systems are increasingly used in community corrections to enforce spatial, temporal, and behavioral rules through continuous sensing. While prior work has examined EM as a criminal justice tool or as a mechanism for compliance, less is known about how sensed data become meaningful in everyday practice. This poster examines EM as a dual-sided sensing system in which supervised individuals and authorities reason about the same data stream from different positions. Based on semi-structured interviews with 26 supervised individuals and 12 authorities in China’s community corrections system, we show that supervised individuals infer system logic from outcomes with limited visibility into how data are interpreted, while authorities reconstruct behavior from ambiguous traces using contextual knowledge, professional experience, and institutional procedures. We call this structural divergence \emph{interpretive misalignment}. It emerges from asymmetric access to data, context, and reasoning processes, and it shapes behavior through probing, strategic adaptation, over-compliance, disengagement, and contestation. We contribute a CSCW account of continuous sensing as distributed interpretive work and identify design opportunities for making data-to-decision processes more legible, contestable, and accountable across system sides.
\end{abstract}


\maketitle

\section{Introduction}
Electronic monitoring (EM) systems use continuous sensing to supervise people under community-based correctional arrangements, translating everyday movement and activity through GPS ankle monitors, mobile reporting tools, geofences, and device-status logs into records that may trigger warnings, intervention, or sanction. We argue that EM should be understood not only as a criminal justice technology, but also as an intensified and highly consequential form of ubiquitous computing, where sensing is persistent, participation is compulsory, and interpretation can directly shape freedom, mobility, work, and family life \cite{weiser1991computer,abowd2000charting,belur2020systematic,zhou2026srl}.

However, the traces produced by EM systems do not carry stable meanings on their own. A boundary crossing may be read as an intentional violation, but may also result from GPS drift, environmental interference, or an imprecisely defined legal boundary; similarly, a signal interruption may indicate device failure, a signal blind spot, or deliberate manipulation \cite{hightower2001location,lane2010survey,troshynski2008accountabilities,zhang2026pervasive}. What becomes consequential, therefore, is not only whether sensing is technically accurate, but how data traces are interpreted, contextualised, and connected to institutional action.

This study examines how different actors construct meaning around the same continuous data stream. Supervised individuals typically begin from their own lived actions and try to infer how the system records, classifies, and judges them, whereas authorities begin from system-generated traces and work backward to reconstruct behaviour. These actors therefore reason in opposite directions, under unequal access to context, discretion, and each other’s interpretive processes. We investigate this dynamic through 38 semi-structured interviews in China’s community corrections system, where EM is used to supervise people serving non-custodial sentences.

This poster contributes a dual-sided empirical account of how continuous sensing is understood by both monitored individuals and institutional reviewers, develops \emph{interpretive misalignment} as a concept for explaining how actors can reasonably draw different meanings from the same data, and identifies design implications for CSCW systems that must support not only sensing accuracy, but also interpretive legibility, contestability, and accountability.

\section{Related Work and Positioning}

\subsection{Electronic Monitoring as Continuous Sensing}
Electronic monitoring has often been evaluated through policy questions such as deterrence, compliance, recidivism, and alternatives to incarceration \cite{belur2020systematic,nellis2014understanding,cao2026causalinfluencemaximizationsteadystate,meng2026focus}. CSCW and HCI research, however, invite a complementary view: EM can be understood as a continuous sensing environment in which data infrastructures reorganize everyday action, responsibility, and institutional interaction. Location-based systems already raise questions about what location records can prove, who is accountable for their interpretation, and how presence becomes administratively meaningful \cite{troshynski2008accountabilities,dourish2004context,su2025flymethrough,su2026capnav}. EM makes these questions especially visible because participation is mandatory and the consequences of interpretation are immediate and high stakes.

\subsection{Sensemaking Around Opaque Systems}
Prior work on algorithmic opacity shows that users develop folk theories and practical models for systems whose internal logic is only partially visible \cite{rader2015beliefs,eslami2016folk,kizilcec2016transparency,chen2026not}. Such work typically focuses on how users explain feeds, rankings, or algorithmic recommendations. In EM, sensemaking occurs under stronger constraints: users cannot simply opt out, and uncertain interpretations may produce sanctions. The system therefore becomes something users must learn, test, and anticipate in daily life.

\subsection{Institutional Judgment and Distributed Interpretation}
Authorities also engage in sensemaking. Street-level bureaucracy scholarship shows that frontline actors apply discretion when formal rules meet incomplete information \cite{lipsky1980street,cao2026beyond,liu2025supporting,ma2026can}. CSCW and algorithmic accountability research further shows that human judgment is shaped by how data, categories, and decision-support systems organize work \cite{green2019principles,alkhatib2019street,bowker1999sorting,luo2025s}. Our study connects these perspectives by examining user-side and authority-side interpretation together. The central problem is not simply opacity for users or discretion for institutions, but how sensing systems distribute interpretive authority across actors.

\section{Method}

We conducted a qualitative interview study with two participant groups in China’s community corrections system: 26 supervised individuals and 12 authorities. Supervised participants had at least three months of recent EM experience, while authorities had direct experience reviewing monitoring records, interpreting anomalies, or making intervention decisions. Participants were recruited through purposive and snowball sampling via community corrections channels, and varied in age, gender, education, occupation, and urban--rural background. Most supervised participants wore GPS ankle monitors; some also used mobile reporting tools, smart wristbands, alcohol sensors, or physiological sensors.

Interviews lasted approximately 45--65 minutes and were conducted one-on-one, online or offline. Interviews with supervised individuals focused on daily restrictions, perceived system logic, mismatched records, uncertainty, coping strategies, and perceptions of authority judgment. To move beyond general attitudes, we asked participants to reconstruct concrete events, such as unexpected system records, moments of uncertainty about whether an action would be flagged, or situations where monitoring changed their routines. Interviews with authorities focused on data presentation, anomaly interpretation, contextual information, uncertainty management, risk assessment, and decisions about whether to observe, verify, warn, or escalate.

We analyzed transcripts using iterative thematic analysis \cite{braun2006using,zhang2025slideaudit,zhang2025openhoiopenworldhandobjectinteraction,chen2025gestobrush}. Three researchers first independently coded a subset of transcripts and developed an initial codebook, including codes such as ``GPS drift,'' ``hidden threshold,'' ``self-protection,'' ``trajectory-based judgment,'' ``risk category,'' and ``evidence conflict.'' We then refined and applied the codebook across the full dataset, using constant comparison \cite{glaser2017discovery,meng2026creating,zhao2025immersive,meng2026decoration} to identify patterns within and across the two participant groups. The analysis focused on how actors interpreted data, how their interpretations diverged, and how such divergence shaped action over time.

\section{Results}
Our findings identify a dual-sided interpretive process. Supervised individuals and authorities make sense of the same monitoring system, but they do so through different starting points, different evidence, and different forms of visibility.

\subsection{Supervised Individuals Build Practical Models of the System}
For supervised individuals, EM was experienced as a set of spatial, temporal, device-related, and behavioral constraints. Participants knew many formal rules, such as not leaving a specified area or returning home before a certain time, but they often did not know how the system translated action into data or how authorities would interpret borderline cases. As a result, they built practical models of the system through experience.

This sensemaking often began when official rules failed to explain system outcomes. Participants described cases in which GPS drift made them appear outside a boundary, arrival time depended on where the system recognized ``home,'' or device power loss was treated as a serious event. These mismatches encouraged participants to infer hidden thresholds: how far GPS might drift, how early they should return, how often a device must be charged, and which behaviors were likely to be detected.

Participants then tested these models through low-risk probing. Some stood near geofence edges to observe location drift, slightly adjusted check-in times, or monitored battery life to estimate device tolerance. These acts were not simply attempts to break rules. Rather, they were ways of making an opaque system learnable. Participants used small experiments to answer practical questions: Where does the boundary really begin? How much delay is tolerated? Does one low-battery event matter? Which device signals are noticed by authorities?

Over time, probing produced more stable behavioral strategies. Some participants used strategic adaptation, such as drawing personal safety maps, setting phone reminders, avoiding known signal blind spots, or keeping informal records of their own movements. Others adopted conservative compliance, such as avoiding boundary areas entirely, returning much earlier than required, charging devices before they were low, or documenting their own location with photos. In both cases, behavior was shaped not only by formal rules, but by uncertainty about how rules would be interpreted.

\subsection{Authorities Reconstruct Behavior from Ambiguous Traces}
Authorities encountered EM primarily through data interfaces: coordinates, trajectories, device states, alerts, and time-stamped anomalies. These traces were compressed and incomplete. A single alert rarely indicated what actually happened. Authorities therefore interpreted data by reconstructing plausible behavior from traces.

Their interpretation relied on contextual judgment. Spatial alerts were assessed in relation to duration, distance, boundary ambiguity, and known GPS problems. Temporal deviations were evaluated through trajectory continuity and whether the person appeared to be returning. Device anomalies were interpreted through environmental knowledge, such as underground garages, dense buildings, or rural signal blind spots. Behavioral restrictions, such as prohibited contact or alcohol use, required even more inferential work because the relevant behavior was rarely directly sensed.

Authorities also developed heuristics through repeated exposure. They compared anomalies against personal history, routine patterns, risk categories, and prior interactions. A one-time delay might be ignored, while repeated minor delays could indicate attitude or risk. A brief signal loss might be treated as technical noise, while frequent offline events could prompt verification. Decision-making under uncertainty often involved staged responses: observing first, calling the person, requesting contextual evidence, or escalating only when patterns accumulated.

\subsection{Interpretive Misalignment Emerges Between the Two Sides}
Although both groups engaged in reasonable sensemaking, their interpretations diverged because the system distributed information unevenly. Supervised individuals reasoned from behavior to data: ``What will the system think I did?'' Authorities reasoned from data to behavior: ``What likely happened behind this trace?'' This opposing direction of inference created different mental models around the same events.

A critical driver of misalignment was invisible interpretive tolerance. Authorities often described flexible practices: brief boundary crossings, small delays, or signal interruptions might be contextualized rather than punished. But this tolerance was rarely visible to supervised individuals. Interfaces typically showed records and outcomes, not the reasoning behind them. Consequently, participants often perceived the system as rigid, automatic, and unforgiving, even when authorities were applying discretionary judgment.

This misalignment had several consequences. Some participants became overly cautious, shrinking their mobility and avoiding ordinary activities. Some engaged in strategic probing to learn hidden thresholds. Others disengaged when uncertainty felt unmanageable or when sanctions seemed arbitrary. In more contentious cases, participants challenged the legitimacy of system records, while authorities continued to treat records and trajectory patterns as the best available evidence. Conflict therefore arose not only over what happened, but over what should count as evidence.

\section{Discussion}

\subsection{Continuous Sensing as Distributed Interpretive Work}
This study reframes EM as a CSCW problem: the system coordinates work across actors who must interpret, verify, and act on partial representations of behavior. Sensing does not simply automate observation; it redistributes interpretive work. Supervised individuals do anticipatory work to make themselves legible to the system. Authorities do reconstructive work to translate traces into decisions. Misalignment arises when these two forms of work remain structurally separated.

This perspective extends prior research on opaque systems by moving from individual understanding to cross-actor coordination. In high-stakes sensing systems, the question is not only whether users understand the system, but whether different actors can align their interpretations of data, context, and evidence. Interpretive misalignment is therefore a relational and organizational condition, not merely a cognitive deficit.

\subsection{From Sensing Accuracy to Interpretive Accountability}
Our findings also suggest that system reliability has at least two layers. The first layer is sensing accuracy: whether the system correctly records location, time, device state, or other signals. The second layer is interpretive accuracy: whether those records are correctly translated into behavioral judgments. A system can fail at either layer. Even accurate data may be misread without context, while inaccurate data can amplify uncertainty at the interpretive layer.

This two-layer structure complicates transparency. Showing raw data is not enough, because raw data may still be ambiguous. Conversely, showing decisions without explaining the interpretive path can make discretionary judgment appear mechanical. For high-stakes sensing systems, accountability requires visibility into both how data were produced and how they were interpreted \cite{ananny2018seeing,selbst2019fairness,meng2026engagement,meng2026tibetcpr}.

\subsection{Design Implications for Dual-Sided Sensing Systems}
We identify \emph{interpretive legibility} as a design principle for dual-sided sensing systems. Systems should make not only sensed records, but also the uncertainty, context, and reasoning that connect records to decisions visible across actors.

First, systems should represent uncertainty directly. Location records might include confidence ranges, drift warnings, and annotations for known signal blind spots. Alerts could distinguish brief boundary noise from sustained departures. Such designs would help both users and reviewers avoid treating all traces as equally authoritative.

Second, systems should support evidence translation. Supervised individuals already produced informal evidence, such as photos or time records, to protect themselves. Interfaces could formalize this practice by allowing timestamped contextual submissions, explanations, or third-party confirmations that authorities can evaluate alongside system traces. This would shift users from risky boundary probing toward safer forms of contextual communication.

Third, contestation should be built into ordinary interpretation rather than reserved for post-sanction appeal. Many conflicts arose because users and authorities relied on incompatible evidentiary frames. Systems should therefore allow users to see why a record matters, what assumptions were applied, and how they can add context before decisions solidify.

Fourth, reviewer tools should support accountable discretion. Authorities need triage mechanisms, consistency checks, anomaly histories, and auditable intervention logs. These features should not remove human judgment; they should make judgment more consistent, reviewable, and explainable \cite{parasuraman1997humans,green2019principles}. In particular, systems could record why an alert was ignored, verified, or escalated, making invisible interpretive tolerance visible without eliminating necessary flexibility.

\subsection{Limitations and Future Work}
This study is situated in China’s community corrections system, where legal, institutional, and cultural conditions shape both supervision and contestation. The findings should therefore be extended carefully to other jurisdictions and domains. Our data also rely on retrospective interviews rather than direct observation of real-time system use. Future work could combine interviews with diary studies, log analysis, or in-situ observation to examine how interpretive models evolve around specific system events. Comparative studies across workplace monitoring, insurance telematics, eldercare sensing, and healthcare monitoring could further test how interpretive misalignment varies by stakes, voluntariness, and accountability structures.

\section{Conclusion}
Electronic monitoring shows that continuous sensing systems govern not only by collecting data, but by organizing how data are interpreted. Supervised individuals and authorities both made sense of the same data stream, but they did so from different positions and with unequal visibility. This produced interpretive misalignment, shaping cautious compliance, probing, disengagement, and conflict. CSCW research and design should therefore attend not only to sensing accuracy, but to the distributed interpretive work through which data become decisions. Making interpretation legible, contestable, and accountable is essential for more trustworthy high-stakes sensing systems.

\section*{Generative AI Disclosure}

No generative AI or AI-assisted technologies were used in the preparation of this manuscript.

\bibliographystyle{ACM-Reference-Format}
\bibliography{references}

@book{lipsky1980street,
  title={Street-Level Bureaucracy: Dilemmas of the Individual 
         in Public Services},
  author={Lipsky, Michael},
  year={1980},
  publisher={Russell Sage Foundation},
  address={New York},
  url={http://www.jstor.org/stable/10.7758/9781610447713}
}

@article{braun2006using,
  author    = {Braun, Virginia and Clarke, Victoria},
  title     = {Using thematic analysis in psychology},
  journal   = {Qualitative Research in Psychology},
  year      = {2006},
  volume    = {3},
  number    = {2},
  pages     = {77--101},
  doi       = {10.1191/1478088706qp063oa}
}

@book{glaser2017discovery,
  title={Discovery of grounded theory: Strategies for qualitative research},
  author={Glaser, Barney and Strauss, Anselm},
  year={2017},
  publisher={Routledge}
}

@article{abowd2000charting,
  title={Charting past, present, and future research in ubiquitous computing},
  author={Abowd, Gregory D and Mynatt, Elizabeth D},
  journal={ACM Transactions on Computer-Human Interaction (TOCHI)},
  volume={7},
  number={1},
  pages={29--58},
  year={2000},
  publisher={ACM New York, NY, USA}
}

@inproceedings{selbst2019fairness,
  title={Fairness and abstraction in sociotechnical systems},
  author={Selbst, Andrew D and Boyd, Danah and Friedler, Sorelle A and Venkatasubramanian, Suresh and Vertesi, Janet},
  booktitle={Proceedings of the conference on fairness, accountability, and transparency},
  pages={59--68},
  year={2019}
}

@article{belur2020systematic,
  title={A systematic review of the effectiveness of the electronic monitoring of offenders},
  author={Belur, Jyoti and Thornton, Amy and Tompson, Lisa and Manning, Matthew and Sidebottom, Aiden and Bowers, Kate},
  journal={Journal of Criminal Justice},
  volume={68},
  pages={101686},
  year={2020},
  publisher={Elsevier}
}

@article{ananny2018seeing,
  title={Seeing without knowing: Limitations of the transparency ideal and its application to algorithmic accountability},
  author={Ananny, Mike and Crawford, Kate},
  journal={new media \& society},
  volume={20},
  number={3},
  pages={973--989},
  year={2018},
  publisher={SAGE Publications Sage UK: London, England}
}

@article{lane2010survey,
  title={A survey of mobile phone sensing},
  author={Lane, Nicholas D and Miluzzo, Emiliano and Lu, Hong and Peebles, Daniel and Choudhury, Tanzeem and Campbell, Andrew T},
  journal={IEEE Communications magazine},
  volume={48},
  number={9},
  pages={140--150},
  year={2010},
  publisher={IEEE}
}

@article{nellis2014understanding,
  title={Understanding the electronic monitoring of offenders in Europe: Expansion, regulation and prospects},
  author={Nellis, Mike},
  journal={Crime, Law and Social Change},
  volume={62},
  number={4},
  pages={489--510},
  year={2014},
  publisher={Springer}
}

@inproceedings{zhou2026srl,
  title={SRL Proxemics: Spatial Guidelines for Supernumerary Robotic Limbs in Near-Body Interactions},
  author={Zhou, Hongyu and Fan, Chia-An and Dong, Yihao and Takashita, Shuto and Inami, Masahiko and Sarsenbayeva, Zhanna and Withana, Anusha},
  booktitle={Proceedings of the 2026 CHI Conference on Human Factors in Computing Systems},
  pages={1--21},
  year={2026}
}

@article{meng2026engagement,
  title={Engagement Is Not Transfer: A Withdrawal Study of a Consumer Social Robot with Autistic Children at Home},
  author={Meng, Yibo and Fan, Guangrui and Liu, Bingyi and Sun, Yingfangzhong and Chen, Ruiqi and Mi, Haipeng},
  journal={arXiv preprint arXiv:2604.02642},
  year={2026}
}

@article{meng2026tibetcpr,
  title={TibetCPR: A Multimodal Tactile Feedback System to Enhance Cardiopulmonary Resuscitation Training in High-Altitude Regions of Tibet},
  author={Meng, Yibo and Chen, Ruiqi and Liu, Zhiming and Ding, Xiaolan},
  journal={arXiv preprint arXiv:2606.07765},
  year={2026}
}

@inproceedings{su2026capnav,
  title={CapNav: Benchmarking Vision Language Models on Capability-conditioned Indoor Navigation},
  author={Su, Xia and Chen, Ruiqi and Liu, Benlin and Ma, Jingwei and Di, Zonglin and Krishna, Ranjay and Froehlich, Jon},
  booktitle={Proceedings of the IEEE/CVF Conference on Computer Vision and Pattern Recognition},
  pages={4043--4053},
  year={2026}
}

@inproceedings{luo2025s,
  title={"What's Happening" - A Human-centered Multimodal Interpreter Explaining the Actions of Autonomous Vehicles},
  author={Luo, Xuewen and Ding, Fan and Panda, Rishikesh and Chen, Ruiqi and Loo, Junnyong and Zhang, Shuyun},
  booktitle={Proceedings of the Winter Conference on Applications of Computer Vision},
  pages={1163--1170},
  year={2025}
}

@article{ma2026can,
  title={Can AI Agents Answer Your Data Questions? A Benchmark for Data Agents},
  author={Ma, Ruiying and Shankar, Shreya and Chen, Ruiqi and Lin, Yiming and Zeighami, Sepanta and Ghosh, Rajoshi and Gupta, Abhinav and Gupta, Anushrut and Gopal, Tanmai and Parameswaran, Aditya G},
  journal={arXiv preprint arXiv:2603.20576},
  year={2026}
}

@article{cao2026beyond,
  title={Beyond Agreement: Scoring Panel-Surfaced Biomedical Entity Candidates for Curator Triage},
  author={Cao, Shuheng and Chen, Ruiqi and Cao, Renjie and Zhang, Zhenhao and Zhang, Siyu and Dan, Tingting},
  journal={arXiv preprint arXiv:2605.30826},
  year={2026}
}

@article{liu2025supporting,
  title={Supporting our ai overlords: Redesigning data systems to be agent-first},
  author={Liu, Shu and Ponnapalli, Soujanya and Shankar, Shreya and Zeighami, Sepanta and Zhu, Alan and Agarwal, Shubham and Chen, Ruiqi and Suwito, Samion and Yuan, Shuo and Stoica, Ion and others},
  journal={arXiv preprint arXiv:2509.00997},
  year={2025}
}

@inproceedings{zhang2025slideaudit,
  title={SlideAudit: A Dataset and Taxonomy for Automated Evaluation of Presentation Slides},
  author={Zhang, Zhuohao and Chen, Ruiqi and Zhong, Mingyuan and Wobbrock, Jacob O},
  booktitle={Proceedings of the 38th Annual ACM Symposium on User Interface Software and Technology},
  pages={1--23},
  year={2025}
}

@inproceedings{su2025flymethrough,
  title={FlyMeThrough: Human-AI Collaborative 3D Indoor Mapping with Commodity Drones},
  author={Su, Xia and Chen, Ruiqi and Ma, Jingwei and Li, Chu and Froehlich, Jon E},
  booktitle={Proceedings of the 38th Annual ACM Symposium on User Interface Software and Technology},
  pages={1--14},
  year={2025}
}

@inproceedings{zhao2025immersive,
  title={Immersive Biography: Supporting Intercultural Empathy and Understanding for Displaced Cultural Objects in Virtual Reality},
  author={Zhao, Ke and Chen, Ruiqi and Zhang, Xiaziyu and Wang, Chenxi and Chen, Siling and Wang, Xiaoguang and Wang, Yujue and Tong, Xin},
  booktitle={Proceedings of the 2025 CHI Conference on Human Factors in Computing Systems},
  pages={1--17},
  year={2025}
}

@inproceedings{chen2025gestobrush,
  title={GestoBrush: Facilitating Graffiti Artists' Digital Creation Experiences through Embodied AR Interactions},
  author={Chen, Ruiqi and He, Qingyang and Bao, Hanxi and Choi, Jung and Tong, Xin},
  booktitle={Proceedings of the 18th International Symposium on Visual Information Communication and Interaction},
  pages={1--9},
  year={2025}
}

@article{chen2026not,
  title={"Not Just Me and My To-Do List": Understanding Challenges of Task Management for Adults with ADHD and the Need for AI-Augmented Social Scaffolds},
  author={Chen, Jingruo and Meng, Yibo and Nie, Kexin},
  journal={arXiv preprint arXiv:2603.17258},
  year={2026}
}

@article{zhang2026pervasive,
  title={The Pervasive Blind Spot: Benchmarking VLM Inference Risks on Everyday Personal Videos},
  author={Zhang, Shuning and Li, Zhaoxin and Wen, Changxi and Ma, Ying and Li, Simin and Zhang, Gengrui and Zhang, Ziyi and Meng, Yibo and Zhao, Hantao and Yi, Xin and others},
  journal={Proceedings of the ACM on Interactive, Mobile, Wearable and Ubiquitous Technologies},
  volume={10},
  number={2},
  pages={1--38},
  year={2026},
  publisher={ACM New York, NY, USA}
}

@inproceedings{meng2026focus,
  title={"Focus on social experiences while alive, not who will see me after I die": A study on the tension between end-of-life loneliness and the mechanisms of digital legacy},
  author={Meng, Yibo and Ye, Lyumanshan and Guan, Yue and Mi, Haipeng},
  booktitle={Proc. EUSSET Conf. on Computer-Supported Cooperative Work},
  year={2026},
  organization={EUSSET}
}

@article{meng2026creating,
  title={"Creating the World with Order!": Designing Tangible Toolkit to Support Creative Expression and Wellbeing for Individuals with ASD},
  author={Meng, Yibo and Ye, Lyumanshan and Feng, Yifan and Fu, Rong and Liu, Bingyi and Li, Linghao and Gao, Nan and others},
  year={2026}
}

@article{meng2026decoration,
  title={From Decoration to Co-Presence: Reconfiguring Human--Plant Relations in Urban Workplaces Through Aesthetic Bio-Feedback},
  author={Meng, Yibo and Liu, Bingyi and Li, Siyuan and Huang, Xinran},
  year={2026}
}

@misc{cao2026causalinfluencemaximizationsteadystate,
  title={Causal Influence Maximization with Steady-State Guarantees},
  author={Renjie Cao and Zhuoxin Yan and Xinyan Su and Zhiheng Zhang},
  year={2026},
  eprint={2603.11761},
  archivePrefix={arXiv},
  primaryClass={stat.ME},
  url={https://arxiv.org/abs/2603.11761}
}

@misc{zhang2025openhoiopenworldhandobjectinteraction,
  title={OpenHOI: Open-World Hand-Object Interaction Synthesis with Multimodal Large Language Model},
  author={Zhenhao Zhang and Ye Shi and Lingxiao Yang and Suting Ni and Qi Ye and Jingya Wang},
  year={2025},
  eprint={2505.18947},
  archivePrefix={arXiv},
  primaryClass={cs.CV},
  url={https://arxiv.org/abs/2505.18947}
}

@article{weiser1991computer,
  title={The computer for the 21st century},
  author={Weiser, Mark},
  journal={Scientific American},
  volume={265},
  number={3},
  pages={94--104},
  year={1991}
}

@article{hightower2001location,
  title={Location systems for ubiquitous computing},
  author={Hightower, Jeffrey and Borriello, Gaetano},
  journal={Computer},
  volume={34},
  number={8},
  pages={57--66},
  year={2001}
}

@inproceedings{troshynski2008accountabilities,
  title={Accountabilities of presence: Reframing location-based systems},
  author={Troshynski, Emily and Lee, Charlotte and Dourish, Paul},
  booktitle={Proceedings of the SIGCHI Conference on Human Factors in Computing Systems},
  pages={487--496},
  year={2008}
}

@article{dourish2004context,
  title={What we talk about when we talk about context},
  author={Dourish, Paul},
  journal={Personal and Ubiquitous Computing},
  volume={8},
  number={1},
  pages={19--30},
  year={2004}
}

@inproceedings{rader2015beliefs,
  title={Understanding user beliefs about algorithmic curation in the Facebook news feed},
  author={Rader, Emilee and Gray, Rebecca},
  booktitle={Proceedings of the 33rd Annual ACM Conference on Human Factors in Computing Systems},
  pages={173--182},
  year={2015}
}

@inproceedings{eslami2016folk,
  title={"I always assumed that I wasn't really that close to [her]": Reasoning about invisible algorithms in news feeds},
  author={Eslami, Motahhare and Rickman, Aimee and Vaccaro, Kristen and Aleyasen, Amirhossein and Vuong, Andy and Karahalios, Karrie and Hamilton, Kevin and Sandvig, Christian},
  booktitle={Proceedings of the 2015 CHI Conference on Human Factors in Computing Systems},
  pages={153--162},
  year={2015}
}

@inproceedings{kizilcec2016transparency,
  title={How much information? Effects of transparency on trust in an algorithmic interface},
  author={Kizilcec, Ren{\'e} F.},
  booktitle={Proceedings of the 2016 CHI Conference on Human Factors in Computing Systems},
  pages={2390--2395},
  year={2016}
}

@article{green2019principles,
  title={The principles and limits of algorithm-in-the-loop decision making},
  author={Green, Ben and Chen, Yiling},
  journal={Proceedings of the ACM on Human-Computer Interaction},
  volume={3},
  number={CSCW},
  pages={1--24},
  year={2019}
}

@inproceedings{alkhatib2019street,
  title={Street-level algorithms: A theory at the gaps between policy and decisions},
  author={Alkhatib, Ali and Bernstein, Michael S.},
  booktitle={Proceedings of the 2019 CHI Conference on Human Factors in Computing Systems},
  pages={1--13},
  year={2019}
}

@book{bowker1999sorting,
  title={Sorting Things Out: Classification and Its Consequences},
  author={Bowker, Geoffrey C. and Star, Susan Leigh},
  publisher={MIT Press},
  year={1999}
}

@article{parasuraman1997humans,
  title={Humans and automation: Use, misuse, disuse, abuse},
  author={Parasuraman, Raja and Riley, Victor},
  journal={Human Factors},
  volume={39},
  number={2},
  pages={230--253},
  year={1997}
}

\end{document}